\documentclass[10pt]{elsart}
\usepackage[dvips]{graphicx}
\usepackage{amsmath,amssymb}
\usepackage{url}              
\usepackage{multirow}
\newcommand{\urlBiBTeX}[1]{\url{#1}}   
\journal{arXiv.org (Published, DOI: 10.1016/j.nimb.2006.03.147) }

\begin{document}
\begin{frontmatter}
\title{LibCPIXE: a PIXE simulation open-source library for multilayered samples }

\author[ITN,CMAM]{C. Pascual-Izarra \corauthref{ca}},
\corauth[ca]{Corresponding author: phone +34 91 5616800, fax +34 91 5859413}
\ead{carlos.pascual@iem.cfmac.csic.es}
\ead[url]{http://cpixe.sourceforge.net}
\author[ITN]{N. P. Barradas},
\author[ITN]{M.A. Reis},

\address[ITN]{Instituto Tecnol\'ogico e Nuclear, E.N. 10, 2686-953 Sacav\'em Codex, Portugal}
\address[CMAM]{CMAM, Universidad Aut\'onoma de Madrid, E-28049 Spain}

\begin{abstract}
Most Particle Induced X-ray Emission (PIXE) data analysis codes are not focused on handling multilayered samples. We have developed an open-source library called "LibCPIXE", for PIXE data analysis. It is written in standard C and implements functions for simulating X-ray yields of PIXE spectra taken from arbitrary samples, including multilayered targets. The library is designed to be fast, portable, modular and scalable, as well as to facilitate its incorporation into any existing program. In order to demonstrate the capabilities of the library, a program called CPIXE was developed and used to analyze various real samples involving both bulk and layered samples. Just as the library, the CPIXE source code is freely available under the General Public License. We demonstrate that it runs both under GNU/Linux systems as well as under MS Windows. There is in principle no limitation to port it to other platforms. 
\end{abstract}
\begin{small}
Note: This is a preprint of a work published in Nuclear Instruments and Methods in Physics Research Section B 249, 820-822 (2006). Please cite the published one.
\end{small}

\end{frontmatter}

\section{Introduction}
The PIXE technique constitutes an excellent tool for quantitative characterization of samples. This has been traditionally done with the aid of computer simulation and fitting programs which are able to solve the \emph{inverse} problem of determining the composition of the sample from an iterative fitting of a PIXE spectrum. In the currently available codes \cite{IAEA-PIXE}, the fitting routines pose limitations on the type of samples that can be analyzed in order to guarantee a reasonably non-ambiguous solution. This is the case, \emph{e.g.} in GUPIX, where layered samples can not contain the same element in more than one layer \cite{GUPIX95}. Such a situation is unsatisfactory, specially in fields like materials science where many interesting samples require a more flexible definition.

In order to address this problem, we decided to make use of more advanced techniques for inverse problem solving \cite{NDF-CPIXE} for which we needed a way of simulating the X-Ray yields from a given arbitrary sample under general experimental conditions. In this work we present a collection of computer routines (the "LibCPIXE" library) that can be used to perform such simulations.
The models and databases used in LibCPIXE (based on those from the DATTPIXE code \cite{DATTPIXE92}) are discussed in the next section and the computational details are treated in section 3. After that, some examples are presented of the performance of the library by comparing simulations and selected experimental results.

\section{Simulation model}


The primary objective of the LibCPIXE library is to calculate the characteristic X-Ray emissions produced by the irradiation of a sample with an ion beam.  The number of detected characteristic X-rays, $N_{ij}$, emitted by an element, $i$, due to an electronic transition, $j$, can be written for a layered sample as:


\begin{equation}
N_{ij}=\frac{\Omega}{4\pi}\epsilon_j N \sum_{l}{\left[ T_{jl}X_{il} \int^{E_{l,\text{in}}}_{E_{l,\text{out}}} \frac{\tau_{lj}\left( x\left(E\right)\right)}{S_l\left(E\right)} \sigma_{ij}(E)dE  \right] }
\label{eq:Yield}~,
\end{equation}


where the $l$ index accounts for each layer in the sample, $\Omega$ is the detection solid angle, $\epsilon_j$ is the efficiency of the detector for the energy of the $j$ emission, $N$ is the number of beam particles and $X_{il}$ is the concentration of the $i$ element in the layer $l$. $T_{jl}$ is the transmittance for the $j$ X-ray energy due to absorbers located between the surface of the layer $l$ and the active area of the detector while $\tau_{lj}(x)$ accounts for the transmittance between the position, $x$, inside the $l$ layer and its surface. Finally, $S_l$ is the stopping force for the beam particles in the layer $l$, and $\sigma_{ij}$ is the X-ray production cross section. Note that $S_l$, $\sigma_{ij}$  and $x$ depend on the energy of the particle, $E$ and that the integral covers the range between $E_{l,\text{in}}$ and $E_{l,\text{out}}$, the energies of the particles when entering and exiting the $l$ layer, respectively.

In the current LibCPIXE code, the X-ray production cross sections for protons are calculated using the ionization cross sections from the ECPSSR theory \cite{Brandt81} with semiempirical corrections from Paul \cite{Paul84} and Reis \cite{Reis96b} for the $K$ and $L$ subshells, respectively, in combination with Coster-Kronig fluorescence yields \cite{Krause79} and the Scofield $K$ and $L$ line transition ratios \cite{Scofield74a,Scofield74b}. Ions other than protons are not yet supported but implementing this should not be complicated since previsions were taken when coding this first version.

In order to calculate the transmittances, absorption coefficients for X-rays in the energy range from 1 to 30 keV have been obtained from \cite{Plechaty75}.

As for the stopping forces (\emph{aka} ``stopping powers''), the ZBL85 values \cite{ZBL85} are used by default and Bragg's rule is applied for compound targets. However, if the default values are not precise enough, any other values (including user-defined ones) can be easily incorporated.

LibCPIXE supports arbitrary samples (including thin, bulk, intermediate and layered specimens). In the case of non-thin samples, enhancement of the X-ray yield may occur due to photoionization by the X-rays emitted in any other part of the sample. The treatment of such effect (the so called \emph{secondary fluorescence correction}) is not trivial in the case of multilayered samples. A paper giving all the details about how this is treated in LibCPIXE is under preparation and will be published elsewhere.

The detector efficiency, $\epsilon$, is obtained experimentally by analyzing known samples and using the code to determine $\epsilon_j$ for the main emission lines. The full efficiency curve is then obtained by interpolation on these experimental values.

It must be noted that, currently, LibCPIXE is focused on calculating the yields of the relevant lines of a PIXE spectrum but does not try to actually simulate a real spectrum and, hence, background is not calculated and no convolutions to simulate finite detector resolutions are made. 
In order to compare the simulation results with real data, the experimental spectra can be processed to eliminate background, identify the peaks an extract their areas. Work is in progress to incorporate into LibCPIXE a tool for automating this process, but meanwhile, the publicly available code QXAS \cite{web:QXAS} can be used to that purpose.

%

\section{Computational details}
The LibCPIXE library, which was originally based on the DATTPIXE v5.3 code \cite{DATTPIXE92}, is written in standard C. A Fortran 90 compatibility layer has also been developed so that existing Fortran programs can be linked against LibCPIXE too.

Routines are included in LibCPIXE to facilitate the interfacing: sample and experimental parameter descriptions can be read from ASCII files or passed directly as function parameters. Similarly, results can be directly returned to the caller program and/or be written in ASCII files. The samples and the absorbers are defined in a layer-by-layer basis using either atomic or mass areal densities for the thicknesses.

To build a fast library has been a priority and this was achieved by using several coding optimization techniques. In the first place, all the databases are pre-cached and stopping force interpolation tables are built to save computation time. The  thickness of the internal sub-layers required for numerical integration of eq. \ref{eq:Yield} is not taken to be constant but is dynamically adjusted to guarantee small enough variations of  $\tau_{lj}$,  $S_l$ and $\sigma_{ij}$. Finally, dynamic memory allocation and intensive use of C pointers are done to improve efficiency.

The LibCPIXE library is easily modifiable and scalable thanks to its modular design and the fact of being open source (it is freely available under the GPL license \cite{web:GPL} from \url{http://cpixe.sourceforge.net}). A simple program called CPIXE has been developed to test and demonstrate the capabilities of the library. CPIXE links against LibCPIXE and has been compiled and tested under both GNU/Linux and MS Windows (porting to other systems/platforms should not pose difficulties).

\section{Examples with the CPIXE demonstration code}
The CPIXE program has been used to simulate three example cases for which experimental spectra were acquired
. The spectra used for these examples are: a) 1090 MeV  H$^+$ beam on an pure Au bulk target, b) same beam on a stoichiometric Fe$_2$O$_3$ bulk sample and c)1220 MeV H$^+$ beam on a Mn$_4$Ir thin film ($562\times10^{15} \text{at/cm}^2$) deposited over a Si substrate. The simulated and experimental yields are compared in table \ref{t:results}. A further demonstration of the practical capabilities of the LibCPIXE (beyond the limits of the CPIXE simple example code) is presented in \cite{NDF-CPIXE}.

\begin{table}[hbtp]
\begin{center}
\begin{tabular}{clrr}
Sample & Line & Simulation & Experiment\\
 \hline
\rule{0pt}{3ex}
\multirow{8}*{Au}
	& Au--$L_{\alpha_{1,2}}$	& 7720	& 7874 \\
	& Au--$L_{\beta_{1}}$		& 1880	& 2535 \\
	& Au--$L_{\beta_{2}}$		& 1546	& 1470 \\
	& Au--$L_{\gamma_{1}}$		& 323	& 499 \\
	& Au--$L_{\ell}$		& 325	& 394 \\
	& Au--$L_{\beta_{3}}$		& 155	& 186 \\
	& Au--$L_{\eta}$		& 46	& 69 \\
	& Au--$L_{\gamma_{3}}$		& 51	& 50 \\
\hline
\multirow{2}*{Fe$_2$O$_3$}
	& Fe--$K_{\alpha_{1,2}}$	& $1.04\times 10^5$	& $1.09\times 10^5$ \\
	& Fe--$K_{\beta_{1}}$		& 15079	& 14891 \\
\hline
\multirow{9}*{Mn$_4$Ir/Si}
	 & Si--$K_{\alpha_{1,2}}$ 	& $1.76\times 10^6$	& $1.77\times 10^6$ \\
	 & Mn--$K_{\alpha_{1,2}}$ 	& 41158	& 41473 \\
	 & Mn--$K_{\beta_{1}}$		& 5929	& 6027 \\
	 & Ir--$L_{\alpha_{1,2}}$	& 2511	& 2508 \\
	 & Ir--$L_{\beta_{1}}$		& 634	& 665 \\
	 & Ir--$L_{\beta_{2}}$		& 423	& 460 \\
	 & Ir--$L_{\gamma_{1}}$		& 127	& 128 \\
	 & Ir--$L_{\ell}$		& 119	& 122 \\
	 & Ir--$L_{\eta}$		& 17	& 18 \\

\end{tabular}
\caption{Comparison of simulated and experimental X-ray yields.}
\label{t:results}
\end{center}
\end{table}

From table \ref{t:results} it can be noted that the experimental results ---deconvoluted from the raw spectra using QXAS--- are reasonably well reproduced by the simulations, specially the $K_{\alpha_{1,2}}$ and $L_{\alpha_{1,2}}$ lines. One would typically choose these lines for a fitting procedure in the case that LibCPIXE is integrated into an automated fitting program (as in \cite{NDF-CPIXE}). 

\section{Conclusions}
LibCPIXE is, to our knowledge, the first truly free \cite{web:GPL} and open-source library to be made publicly available to the IBA community for PIXE simulations. It is our aim that developers of existing IBA simulation/analysis codes may consider to make use of this library to extend their codes into supporting the PIXE technique too. LibCPIXE is already functional ---as demonstrated in this paper and in \cite{NDF-CPIXE}--- but more capabilities and extensions are being developed at the moment of writing this.

\section*{Acknowledgments}
Thanks are due to P. C. Chaves and V. Corregidor for their help with the experimental data acquisition. This work has been supported by an EU grant (HPRN-CT-2001-00199).


\begin{thebibliography}{10}
\expandafter\ifx\csname url\endcsname\relax
  \def\url#1{\texttt{#1}}\fi
\expandafter\ifx\csname urlprefix\endcsname\relax\def\urlprefix{URL }\fi

\bibitem{IAEA-PIXE}
S.~Fazini\'c, {Intercomparison of PIXE Spectrometry Software Packages}, TECDOC
  1342, {International Atomic Energy Agency (IAEA)} (2003).

\bibitem{GUPIX95}
J.~A. Maxwell, W.~J. Teesdale, J.~L. Campbell, {The Guelph PIXE software
  package II}, Nucl. Instr. and Meth. B 95 (1995) 407--421.

\bibitem{NDF-CPIXE}
C.~Pascual-Izarra, M.~A. Reis, N.~P. Barradas, {Simultaneous PIXE and RBS data
  analysis using Bayesian Inference with the DataFurnace code}, Nucl. Instr.
  and Meth. B 249, 780-783 (2006).

\bibitem{DATTPIXE92}
M.~A. Reis, L.~C. Alves, {DATTPIXE, a computer package for TTPIXE data
  analysis}, Nucl. Instr. and Meth. B 68 (1992) 300--304.

\bibitem{Brandt81}
W.~Brandt, G.~Lapicki, {Energy-loss effect in inner-shell Coulomb ionization by
  heavy charged particles}, Phys. Rev. A 23 (1981) 1717--1729.

\bibitem{Paul84}
H.~Paul, {An analytical cross-section formula for K X-ray production by
  protons}, Nucl. Instr. and Meth. B 3 (1984) 5--10.

\bibitem{Reis96b}
M.~A. Reis, A.~P. Jesus, {Semi-empirical Approximation Cross Sections for L
  X-ray Production by Proton Impact}, Atomic Data and Nuclear Data Tables 63
  (1996) 1--55.

\bibitem{Krause79}
M.~O. Krause, {Atomic radiative and radiationless yields for K and L shells},
  J. Phys. Chem. Ref. Data 8 (1979) 307--327.

\bibitem{Scofield74a}
J.~H. Scofield, {Exchange Corrections of K x-ray emission rates}, Phys. Rev. A
  9 (1974) 1041--1049.

\bibitem{Scofield74b}
J.~H. Scofield, {Relativitic Hartree-Slater values for K and L x-ray emission
  rates}, Atomic Data and Nucl. Data Tables 14 (1974) 121--137.

\bibitem{Plechaty75}
E.~F. Plechaty, D.~E. Cullen, R.~J. Howeton, {Tables and Graphs of Photon
  Interaction Cross Section from 1.0 keV to 100 MeV Derived from LLL Evaluated
  Nuclear Data Library}, Report DLC-7, Lawrence Radiation Laboratory,
  Livermore, California (1975).

\bibitem{ZBL85}
J.~F. Ziegler, J.~P. Biersack, U.~Littmark, Stopping and Ranges of ions in
  Solids, Pergamon, New York, 1985.

\bibitem{web:QXAS}
{QXAS Seibersdorf Laboratories, IAEA}, {URL}:
  {\urlBiBTeX{http://www.iaea.org/OurWork/ST/NA/NAAL/pci/ins/xrf/pciXRFdown.ph%
p}}.

\bibitem{web:GPL}
{GNU General Public License}, {URL}:
  {\urlBiBTeX{http://www.gnu.org/licenses/gpl.html}}.

\end{thebibliography}

\end{document}